\documentclass[3p,times,procedia]{elsarticle}
\usepackage{nupha_ecrc}
\usepackage{multirow}

\volume{00}

\firstpage{1}

\journalname{Nuclear Physics A}

\runauth{A. Czajka, S. Jeon}

\jid{nupha}

\jnltitlelogo{Nuclear Physics A}

\usepackage{amssymb}
\usepackage{amsthm}

\usepackage[figuresright]{rotating}

\newcommand{\be}{\begin{equation}}
\newcommand{\ee}{\end{equation}}
\newcommand{\ba}{\begin{eqnarray}}
\newcommand{\ea}{\end{eqnarray}}
\newcommand \nn {\nonumber}

\begin{document}

\begin{frontmatter}




\title{The shear and bulk relaxation times from the general correlation functions}


\author[mg,ujk]{Alina Czajka}
\author[mg]{Sangyong Jeon}

\address[mg]{Department of Physics, McGill University, 3600 rue University,
Montreal, Quebec H3A 2T8, Canada}
\address[ujk]{Institute of Physics, Jan Kochanowski University, ul. Swietokrzyska 15, 
25-406 Kielce, Poland}

\begin{abstract}
In this paper we present two quantum field theoretical analyses on the shear
and bulk relaxation times. First, we discuss how to find Kubo formulas for
the shear and the bulk relaxation times. Next, we provide 
results on the shear viscosity relaxation time obtained within the diagrammatic approach 
for the massless $\lambda\phi^4$ theory.
\end{abstract}

\begin{keyword}
Kubo formula, shear relaxation time, bulk relaxation time

\end{keyword}

\end{frontmatter}


\section{Introduction}
\label{intro}

Over the course of last decades relativistic viscous hydrodynamics has been shown
to successfully describe and explain the behavior of the strongly interacting matter 
produced in heavy-ion collisions at the
Relativistic Heavy Ion Collider (RHIC) and the Large Hadron Collider (LHC),
see Refs.~\cite{Gale:2013da,Heinz:2013th} and references therein. In general, any viscous 
fluid is characterized by a set of transport coefficients. These enter hydrodynamic
equations as parameters and must be obtained from the underlying microscopic theory, either in
quantum field theory or kinetic theory. Many approaches have been developed to study the first-order transport coefficients \cite{Jeon:1992kk}-\cite{Arnold:2002zm} for various systems.
There are also some studies on second-order transport coefficients \cite{York:2008rr}-\cite{Hong:2010at} but their quantum-field-theoretical determination does not seem to be complete.

We have already undertaken the task to study the shear and the bulk relaxation times consistently 
from first principles. The comprehensive analysis is shown in \cite{Czajka:2017bod} and here only a brief summary is presented. Using general properties of Green functions and the gravitational Ward identity we first parametrize the stress-energy correlation functions to find their most general forms. Then Kubo formulas for the relaxation times are found in the hydrodynamic limits of the corresponding response functions. We also study shear effects in the massless scalar field theory 
$\lambda \phi^4$ and calculate the shear relaxation time within the real-time formalism.

\section{Equations of viscous hydrodynamics}
\label{hydro}

The behavior of a relativistic system, which is close to thermal equilibrium, can be well described
by the viscous hydrodynamics, which is based on the energy-momentum conservation law
\ba
\label{en-mom-conservation}
\partial_\mu T^{\mu\nu}&=&0,
\ea
where the energy-momentum tensor takes the form
\ba
\label{vis-T}
T^{\mu\nu} = \epsilon u^\mu u^\nu - \Delta^{\mu\nu}(P+\Pi) +
\pi^{\mu\nu}
\ea
with $\epsilon$ being the energy density, $P$  - thermodynamic pressure, $u^\mu$ are
the components of the flow velocity with the normalization condition
$u^\mu u_\mu=1$, $\Delta^{\mu\nu} = g^{\mu\nu} - u^\mu u^\nu$ is the
projection operator with $u_\mu \Delta^{\mu\nu} = 0$,
and the Minkowski metric is 
$g^{\mu\nu}=(1,-1,-1,-1)$. The terms
$\Pi$ and $\pi^{\mu\nu}$ are the bulk viscous pressure and the shear
stress tensor, respectively, which have well defined forms in the Navier-Stokes limit. Then the viscous 
corrections are characterized by the bulk viscosity $\zeta$ and shear viscosity $\eta$, respectively.
In the second order formulation of viscous hydrodynamics, the response of medium to the thermodynamic forces is not instantaneous. The viscous corrections approach their corresponding Navier-Stokes forms within some characteristic time scales, which
are the bulk and shear relaxation times, $\tau_\Pi$ and $\tau_\pi$. Consequently, the viscous corrections are subject to relaxation equations
\ba
\Pi=\Pi_{NS} - \tau_\Pi \dot \Pi, \qquad\qquad\qquad
\pi^{\mu\nu}=\pi^{\mu\nu}_{NS} - \tau_\pi \dot \pi^{\langle \mu\nu
\rangle}, 
\ea
where $\Pi_{NS}$ and $\pi^{\mu\nu}_{NS}$ are the bulk pressure and the stress tensor in the Navier-Stokes approach, and we used the notation $A^{\langle \mu\nu \rangle} \equiv
\Delta^{\mu\nu}_{\alpha\beta}A^{\alpha\beta}$ where $\Delta^{\mu\nu}_{\alpha\beta} \equiv (\Delta^\mu_\alpha
\Delta^\nu_\beta +
\Delta^\mu_\beta \Delta^\nu_\alpha
-2/3\Delta^{\mu\nu}\Delta_{\alpha\beta})/2$. We will not consider the non-linear terms here. For more advanced studies on the hydrodynamic equations see, for example, \cite{Jeon:2015dfa}.

In case when there are no other than energy and momentum currents occurring in the system, two
hydrodynamic modes determine its dynamics. They are governed by the following dispersion relations
\ba
\label{disp-rel-mom}
0 &=& -\omega^2 \tau_\pi -i \omega +D_T {\bf k}^2 \\
\label{disp-rel-en}
0&=& - \omega^2 + v_s^2 {\bf k}^2
+ i\omega^3 (\tau_\pi + \tau_\Pi)
- i\bigg(\frac{4D_T}{3}+\gamma+v_s^2(\tau_\pi + \tau_\Pi) \bigg)\omega{\bf k}^2 
\\ \nn
&&
+\tau_\pi \tau_\Pi \omega^4 - \tau_\pi \tau_\Pi v_s^2 \omega^2 {\bf k}^2
- \tau_\Pi \frac{4D_T}{3} \omega^2 {\bf k}^2 
- \tau_\pi \gamma \omega^2 {\bf k}^2 ,
\ea
where $\omega$ and ${\bf k}$ are the frequency and wavevector of the modes, $D_T=\eta/(\epsilon+P)$, $\gamma=\zeta/(\epsilon+P)$, and $v_s^2=\partial P/\partial \epsilon$ is the speed of sound squared. The dispersion relation (\ref{disp-rel-mom}) governs the propagation of the diffusion mode which occurs in the direction transverse to the the flow velocity. The sound mode is, in turn, given by 
the dispersion relation (\ref{disp-rel-en}) and it is an effect of the small disturbances propagating 
longitudinally in the medium. Both dispersion relations are essential to determine the respective correlation functions of the stress-energy tensor.

\section{Stress-energy correlation functions and Kubo formulas}
\label{kubo}

Since viscous hydrodynamics is a manifestation of the linear response theory, the deviations of 
different observables are given in terms of corresponding equilibrium response functions. Therefore,
the response functions carry dynamical information about the system. In general, these functions 
cannot be calculated exactly but one is able to parametrize their most general structures for the stress-energy tensor components using the following arguments. First, the real part of a correlation function is an even function of frequency and the imaginary part, which directly corresponds to the spectral function, must be an odd function of frequency. Next, since the stress-energy tensor represents at the same time the conserved currents and also generators of the space time evolution, its correlation functions
must satisfy the gravitational Ward identity \cite{Deser:1967zzf}
\ba
\label{Ward-id-real}
k_\alpha \big( \bar G^{\alpha\beta,\mu\nu}(k) 
- g^{\beta\mu} \langle \hat T^{\alpha\nu} \rangle 
- g^{\beta\nu} \langle \hat T^{\alpha\mu} \rangle 
+ g^{\alpha\beta} \langle \hat T^{\mu\nu} \rangle \big)=0.
\ea
Finally, the low-frequency and long-wavelength limits must be properly incorporated to ensure the
correlation functions behave well in these limits.

Using aforementioned arguments we have parametrized the retarded Green function for the shear 
flow as
\ba
\label{G-xyxy}
\bar G_R^{xy,xy}(\omega,k_y) = 
\frac{\omega^2(\epsilon + g_T(k_y) + i\omega A(\omega,k_y))}
{k_y^2 - i\omega/D(\omega,k_y)-\omega^2 B(\omega,k_y)} - P,
\ea
where $g_T(k_y) = P + g_{\pi\pi}(k_y)$ with $g_{\pi\pi}(k_y)$ being a function of $k_y$ of the order at least $O(k_y^2)$ in the small $k_y$ limit. The functions $A$, $B$, and $D$ have the form
$D(\omega,k_y)=D_R(\omega,k_y)-i\omega D_I(\omega,k_y)$,
where $D_R(\omega,k_y)$ and $D_I(\omega,k_y)$ are real-valued even
functions of $\omega$ and $k_y$. The real parts $D_R$ and $B_R$ must
have a non-zero limit as $\omega \rightarrow 0$ and $k_y \rightarrow 0$. All
other parts of $A$, $B$, and $D$ must have finite limits as $\omega
\rightarrow 0$ and $k_y \rightarrow 0$. By matching the denominator of the Eq.~(\ref{G-xyxy})
to the dispersion relation (\ref{disp-rel-mom}) one is able to express $D_R$, $D_I$, $B_R$, etc in the hydrodynamic limits via the transport coefficients and obtain the following Kubo relations
\ba
\label{KF-1r}
\eta = \lim_{\omega \rightarrow 0} \lim_{k_y \rightarrow 0}
\frac{1}{\omega} \textrm{Im}\bar
G_R^{xy,xy}(\omega, k_y), \qquad\qquad\qquad
\eta \tau_\pi - \frac{\kappa}{2}=
-\frac{1}{2} \lim_{\omega \rightarrow 0} \lim_{k_y \rightarrow 0}
\partial_\omega^2 \;
\textrm{Re} \bar G_R^{xy,xy}(\omega,k_y),
\ea
where $\kappa$ is the static susceptibility. Accordingly, both formulas have to be used to find the shear relaxation time in the leading order. The coefficient $\kappa$ was shown in \cite{Romatschke:2009ng,Moore:2012tc} to be $\kappa = \mathcal{O}(\lambda^0 T^2)$ in the weak coupling limit.
On the other hand, both $\eta$ and $\tau_\pi$ are related to the mean free path
which depends inversely on the coupling constant. Hence, $\kappa$ may be omitted in further analysis. 

To examine the bulk viscosity and its relaxation time one needs to consider the response function
to the longitudinal fluctuations. Employing the analogous procedure as in case of shear flow
we obtain the most general form of the longitudinal fluctuation correlation function in the form
\ba 
\label{GL-gen-form}
\bar G_L(\omega,{\bf k})=
\frac{\omega^2(\epsilon+ P  +\omega^2 Q(\omega, {\bf k}) )}
{{\bf k}^2 - \omega^2/Z(\omega, {\bf k})
+ i\omega^3R(\omega, {\bf k})},
\ea
where the functions $Q$, $R$, and $Z$ are constrained in a similar way as the functions $A$, $B$, and
$D$ occurring in Eq. (\ref{G-xyxy}). When hydrodynamic limits are applied one obtains some Kubo formulas employing $\zeta$, $\eta$, $\tau_\Pi$, $\tau_\pi$, and $\kappa$. However, with the help of
Eqs. (\ref{KF-1r}) we can extract simple Kubo formulas involving only bulk viscosity and its relaxation time. These are then given in terms of the pressure-pressure correlation function
\ba
\label{Kubo-final-1}
\zeta =
\lim_{\omega \rightarrow 0} \lim_{{\bf k} \rightarrow 0} 
\frac{1}{\omega} \textrm{Im}\,\bar{G}_R^{PP}(\omega, {\bf k}), \qquad\qquad\qquad
\zeta \tau_\Pi =
-\frac{1}{2} \lim_{\omega \rightarrow 0} \lim_{{\bf k} \rightarrow 0} 
\partial^2_\omega \textrm{Re}\, \bar{G}_R^{PP}(\omega, {\bf k}).
\ea
These formulas constitute the way of computation of the bulk relaxation time.

\section{Shear relaxation time in scalar field theory}
\label{shear-time}

It is elastic scatterings that constitutes the shear effects and therefore to evaluate the shear relaxation time within the scalar massless field theory $\lambda \phi^4$ one requires expanding the retarded Green function to the leading order. For the comprehensive calculation see \cite{Czajka:2017bod}.

\begin{figure}[!h]
\centering
\includegraphics*[width=0.35\textwidth]{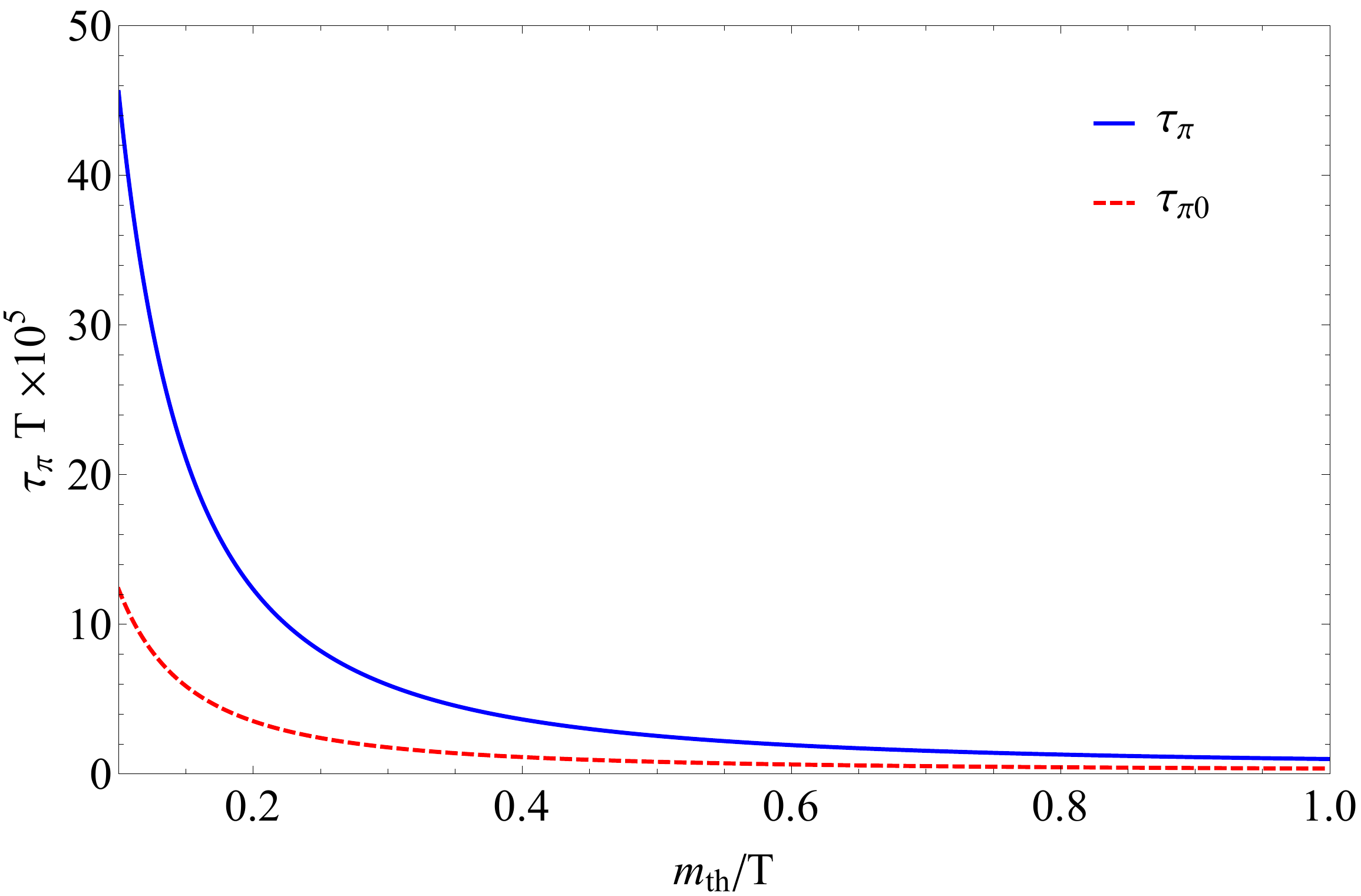}
\hspace{1cm}
\includegraphics*[width=0.35\textwidth]{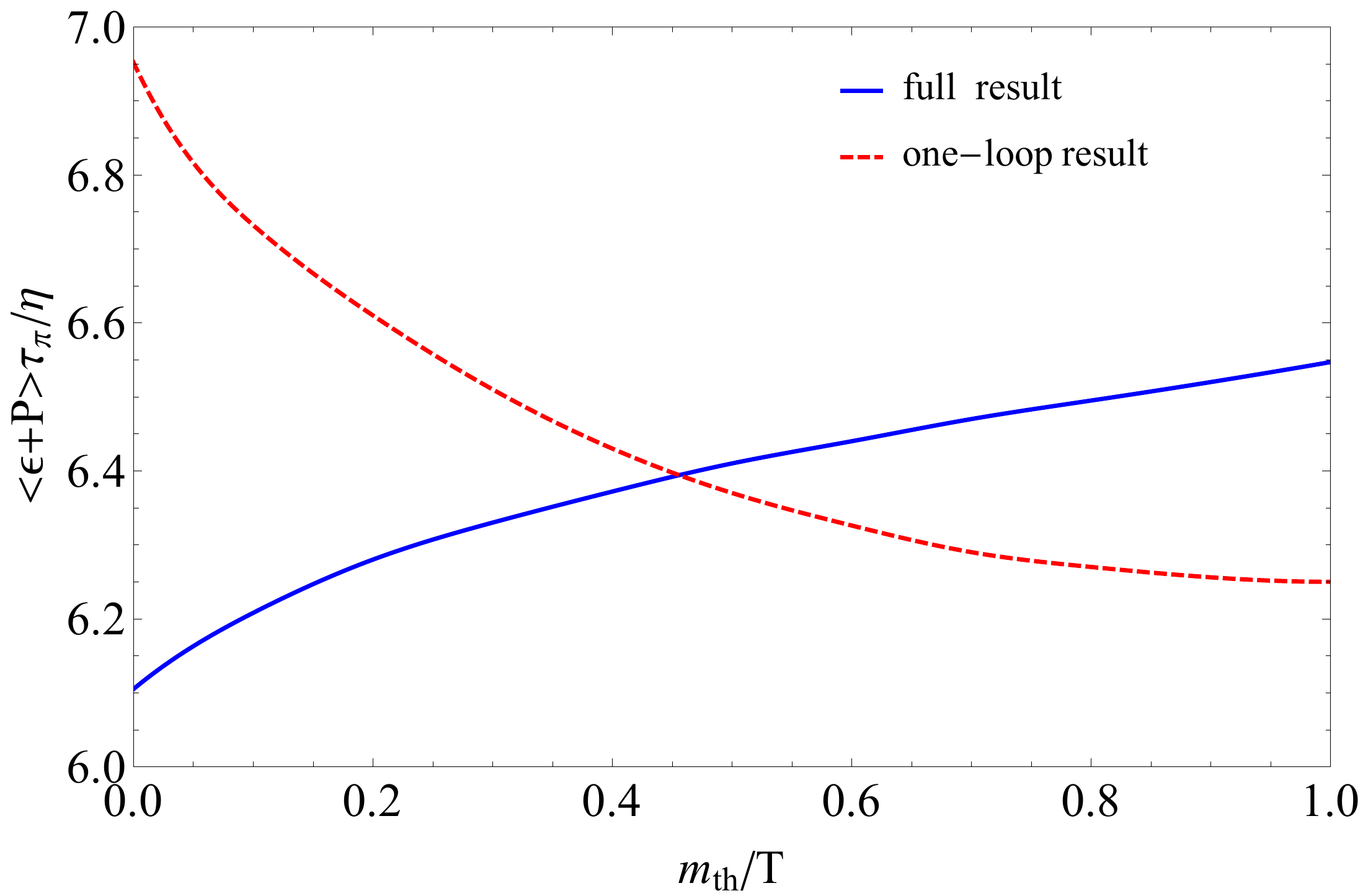}
\caption{Left panel: shear relaxation time as a function of $m_{\textrm {th}}/T$.
Right panel: the ratio $\langle\epsilon+P\rangle\tau_\pi/\eta$ as a function of $m_{\textrm
{th}}/T$. Evaluation in the one-loop limit (red dashed curve) and multi-loop
resummation (blue curve) are presented. }
\label{fig-tau-shear}
\end{figure}

Following Kubo relations one obtains integral equations for the imaginary and real parts of the retarded 
Green functions, which then are solved numerically and the shear relaxation time is extracted. It has been shown that both the shear viscosity and the relaxation time are related to the inverse of the
thermal width, as expected. In the left panel of Fig.~\ref{fig-tau-shear} the relaxation time is
plotted as a function of the thermal mass over temperature and $m_{\textrm{th}}/T$ which should be identified with $\sqrt{\lambda}$ with $\lambda$ being the coupling constant. In the left panel of 
Fig.~\ref{fig-tau-shear} we present the ratio $\langle\epsilon+P\rangle\tau_\pi/\eta$ as a function 
of $m_{\textrm{th}}/T$ showing that it changes between 6.11 up to 6.55 in the range shown. The 
results presented here are in agreement with those obtained within kinetic theory \cite{York:2008rr, Denicol:2014vaa}.

\section{Conclusions}
\label{conclusions}

In this article we presented a set of Kubo formulas which enable us to compute the shear and
the bulk relaxation times. The main focus was to obtain the Kubo formula to compute the bulk relaxation time and the formula found is consistent with the one obtained via the metric perturbation analysis \cite{Hong:2010at} but it is different from the one obtained from the projection operator method \cite{Huang:2010sa}.

We also computed the shear relaxation time and the ratio $\langle\epsilon+P\rangle\tau_\pi/\eta$ within scalar massless theory and their values are consistent with previous studies. In general the ratio
seems to vary between 5 up to 7 across different theories.

\section*{Acknowledgments}

A. Czajka acknowledges the support from the program Mobility Plus of the
Polish Ministry of Science and Higher Education.
S.J. is supported in part by the Natural Sciences and
Engineering Research Council of Canada.
Discussions with C.~Gale and G.D.~Moore are very much appreciated.






\end{document}